\documentclass[pre,twocolumn,superscriptaddress,showkeys,showpacs]{revtex4}
\usepackage{amssymb}
\usepackage[english]{babel}

\newcommand{\kb}{\mathrm{k_B}}
\newcommand{\teff}{T_{\mathrm{eff}}}
\newcommand{\ep}{\varepsilon_{p}}

\begin{document}

\title{Collisional cross sections and momentum distributions in astrophysical plasmas: dynamics and statistical mechanics link}

\author{F.~Ferro}
\email{fabrizio.ferro@polito.it}
\affiliation{Dipartimento di Fisica, Politecnico di Torino, I-10129 Torino, Italy}
\affiliation{INFN - Sezione di Torino, I-10125 Torino, Italy}
\author{P.~Quarati}
\email{piero.quarati@polito.it}
\affiliation{Dipartimento di Fisica, Politecnico di Torino, I-10129 Torino, Italy}
\affiliation{INFN - Sezione di Cagliari, I-09042 Monserrato, Italy}

\begin{abstract}
We show that, in stellar core plasmas, the one-body momentum distribution function is strongly dependent, at least in the high velocity regime, on the microscopic dynamics of ion elastic collisions and therefore on the effective collisional cross sections, if a random force field is present. We take into account two cross sections describing ion-dipole and ion-ion screened interactions. Furthermore we introduce a third unusual cross section, to link statistical distributions and a quantum effect originated by the energy-momentum uncertainty owing to many-body collisions, and propose a possible physical interpretation in terms of a tidal-like force. We show that each collisional cross section gives rise to a slight peculiar correction on the Maxwellian momentum distribution function in a well defined velocity interval. We also find a possible link between microscopical dynamics of ions and statistical mechanics interpreting our results in the framework of non-extensive statistical mechanics.
\end{abstract}

\date{\today}

\pacs{51.10.+y, 95.30.Dr, 05.20.-y}
\keywords{non-extensive statistical mechanics, astrophysical plasmas, many-body collisions, quantum uncertainty}

\maketitle

\section{Introduction}
In a dense neutral plasma like, for instance, the astrophysical plasma of a stellar core with mean
potential energy of about the same order of magnitude as thermal energy~\cite{Clayton1968,Rolfs1988}, a random
microscopical electric field (usually called electric microfield) arises at any spatial point. Its origin does not
rely on correlations between particles in the plasma, and indeed it is present also in ideal plasmas of
statistically independent particles; rather, it is due to local thermal fluctuations in the position of
ions~\cite{Iglesias1983}. Commonly, the microfield strength is not negligible, being of the same order of
magnitude of the Coulomb field of a unit charge at the characteristic Wigner-Seitz radius. As Romanovsky
and Ebeling~\cite{Romanovsky1998} pointed out, the dynamic enhancement on nuclear fusion rates due to electric
random fields is large only in very dense stars like white dwarfs; on the contrary, its importance inside the
Sun's core is presently believed to be limited.

We show that a random field of generic nature (random electric or magnetic microfields belong, among others, to this category) may play a crucial role, as it influences the upper tail of the one-body stationary momentum distribution function of ions in a dense neutral stellar plasma, leading to slight deviations from a pure Maxwellian distribution. Furthermore, different elastic collisional cross sections among interacting ions could influence significantly the tail of the distribution, and each one provides corrections in a characteristic velocity range only.

We want to stress here that, in astrophysical plasmas, many different collisional processes could be active at the
same time, provided that we consider different velocity intervals. Besides, for instance, the usual pure Coulomb interaction (described by the well-known Rutherford cross section formula), many different screening potentials have been proposed since long time, in order to provide effective models for several astrophysical conditions~\cite{screening}. Therefore, screening and many-body effects, whose importance relies on the fact that they strongly enhance thermonuclear reaction rates, are also important, lying on a kinetic framework, as they modify the collisional cross section between ions~\cite{Ferro}.

In this paper we analytically derive the one-body distribution function of momentum starting from a kinetic
equation in which we set three cross sections of interest in dense and weakly non-ideal plasmas. Our calculations are based on the existence of a random force field $F$, which can be justified either by the theory of electric microfields (as outlined in Refs.~\cite{Iglesias1983,Romanovsky1998}), or the theory of dissipative random forces in the approach of Langevin equation for Brownian motion (see, for example Ref.~\cite{Reichl}), and can be originated from density fluctuations in the plasma. We are urged in this line of research also by the Einstein's criticism to the Boltzmann probability relation, based on the argument that statistical mechanics might only be justified in terms of classical or quantum microscopic dynamics~\cite{Einstein}.

In the first part of our paper (Sections~\ref{sect kinetic equation}, \ref{kinetical solution} and~\ref{collisional cross section}) we use a kinetic approach to describe the motion of particles submitted to a generic random force field with finite relaxation time (i.e. not $\delta$-correlated). The stationary solution can be expressed in terms of collisional cross sections and collisional frequencies; therefore we may establish a link between the type of particle collisions and the form of stationary distribution functions that can differ from the equilibrium Maxwellian function. Then, we define the parameter $q$ characterizing the deformation factor of our distribution and we calculate it in terms of known physical quantities. Interpreting this deformation on the physical ground of non-extensive statistical mechanics, and Tsallis statistics as a special case~\cite{nonextensive}, the parameter $q$ can be expressed in terms of dynamical quantities like cross section, ion-correlation parameter and plasma parameter. A dynamical realization of non-extensive statistics and, in general, of superstatistics has been completed by Beck~\cite{Beck2001} by using stochastic differential equations with spatio-temporally fluctuating parameters.

In the second part (Section~\ref{sect quantum uncertainty}), we consider the quantum energy-momentum uncertainty due
to contemporaneous interaction among many particles of plasma. We use the Kadanoff-Baym ansatz~\cite{BaymKadanoff} and follow the approach of Galitski\v{\i} and Yakimets~\cite{Galitskii} that leads to the appearance of power-law tail in the momentum distribution function. This function shows an enhanced tail at high momentum depending on the collision frequency and the collisional cross section. As a consequence, we show a link between non-extensive statistical mechanics and dynamics, by evidentiating the type of microscopic elastic collisions acting among particles and how the collision frequency is related to momentum distributions. Then we investigate the cross section that reproduces the non-extensive distribution function, limiting ourselves, for simplicity, to the case $q>1$, and we show that the corresponding interaction looks like a tidal-like force superimposed on the two-body attraction, giving a collision cross section $\sigma(\varepsilon_p)\sim\sqrt{\varepsilon_p}$, where $\varepsilon_p$ is the relative kinetic energy. The quantum approach that we follow along our discussion relies on equilibrium conditions of the system; therefore our final result is a real equilibrium (not metastable or stationary) distribution function that differs from the equilibrium Maxwell-Boltzmann distribution. The existence of such an unusual equilibrium function is due to the quantum uncertainty as it is shortly described in Section~\ref{sect quantum uncertainty}.

\section{Kinetic equation under a generalized random force}\label{sect kinetic equation}
A kinetic equation describing electrons in a plasma under an external electric field, in which collisions between
electrons and neutral atoms are present, was derived by Chapman and Cowling~\cite{Chapman1970}, by Spitzer~\cite{Spitzer} and by Golant, \v{Z}ilinskij and Sacharov~\cite{Golant}. They express the actual one-body velocity distribution function as a formal series,
\begin{displaymath}
\tilde{f}({\bf v})=f+f_1+f_2+\dots\; ,
\end{displaymath}
where $f=f(v)$ is the isotropic component (with $v=|{\bf v}|$), while $f_1$,$f_2$ and so on describe next
orders of anisotropy induced by the external field. In addition, $f(v)$ may be only a slight perturbation of the
Maxwellian distribution function.

Here, we adopt their equation in order to derive the momentum stationary distribution of ions, but we replace the external electric field with a generalized random force $F$ and we focus on the isotropic function $f$ only. The elastic collisional cross sections that we are considering describe the interaction among ions and among ions and electric dipoles of polarized neutral compounds of the Wigner-Seitz spheres. All these cross sections will be discussed in the next sections.

Thus, considering the plasma component consisting of ions of mass $m$, the kinetic equation reads~\cite{Golant}
\begin{equation}
\pm\frac{2}{3}\frac{F^2}{\mu^2\nu^2}\frac{\textrm{d} f}{\textrm{d} v}+\kappa\left(vf+\frac{\kb T}{\mu}\frac{\textrm{d} f}{\textrm{d} v}\right)=0\;
,\label{kinetic equation}
\end{equation}
where $v$ is the modulus of the relative velocity between two ions, $\mu$ is their reduced mass, $\nu(v)$ is the
collisional frequency, $\kb T$ is the thermal energy and $\kappa$, which is defined as
\begin{displaymath}
\kappa=2\frac{\mu^2}{m^2}\; ,
\end{displaymath}
is the energy transfer coefficient (or an average value of it).

Let us discuss briefly the origin of the double sign in the term containing the random force $F$ in
Eq.~\ref{kinetic equation}. The quantity
\begin{displaymath}
D_{F}=\pm\frac{2}{3}\frac{F^2}{\mu^2\nu^2}\; ,
\end{displaymath}
is the perturbation on the diffusion coefficient of the system due to the force $F$, and the corresponding
particle flux is given by
\begin{displaymath}
J_{F}=\mp D_{F} n\frac{\textrm{d} f}{\textrm{d} v}\; ,
\end{displaymath}
$n$ being the ion particle density of the plasma. If $F$ were an electric microfield, i.e. $F={\mathrm e} E$ (with
${\mathrm e}$ equal to the electric charge of one ion), the corresponding sign would be positive, thus enhancing
the actual diffusivity of the system, while in the opposite case, the total diffusivity would drop. Therefore, we
introduce the double sign since we want to deal with the most general situation, in which either sub-diffusivity or
super-diffusivity may be significant.

The analytical solution of Eq.~\ref{kinetic equation} reads
\begin{equation}
f(v)=f(0)\cdot\exp\left[-\int_{0}^{v}\textrm{d} v'\frac{\mu v'}{\kb T\pm\frac{2}{3}\frac{F^2}{\mu\kappa\nu^2}}\right]\;
,\label{general solution}
\end{equation}
where the constant $f(0)$ should be calculated through the normalization condition
\begin{displaymath}
4\pi\int_{0}^{+\infty}\textrm{d} v' v'^2 f(v')=1\; .
\end{displaymath}

Let us define now a characteristic strength of the generalized random field $F$ as
\begin{displaymath}
F_{C}=\nu\sqrt{\kappa\mu\kb T}\; .
\end{displaymath}

Then, if the condition
\begin{equation}
F^2\ll F_{C}^2\; ,\label{critical condition}
\end{equation}
holds, i.e. if the random force is negligible, Eq.~\ref{general solution} gives the Maxwellian distribution function at temperature $T$. The central point is that, in this case, the Maxwellian function is the
solution of the kinetic equation~\ref{kinetic equation} regardless of any assumption about the collisional
frequency $\nu$ of the plasma.

On the contrary, if the condition in Eq.~\ref{critical condition} fails, the form of the solution $f(v)$ is determined by the explicit dependence of the collisional frequency $\nu$ on relative velocity $v$. The $\nu(v)$
frequency is itself a function of the collisional cross section $\sigma(v)$, as
\begin{equation}
\nu(v)=nv\sigma(v)\; .\label{collisional frequency}
\end{equation}

Thus, in this case, Eq.~\ref{general solution} leads to a Maxwellian provided that we choose the $\sigma(v)=\alpha_0 v^{-1}$ cross section ($\alpha_0$ being a suitable constant), and that we renormalize the temperature of the plasma in the following fashion,
\begin{equation}
\kb\teff=\kb T\pm\frac{2}{3}\frac{F^2}{\kappa\mu n^2\alpha_0^2}\; ,\label{effective temperature}
\end{equation}
where $\teff$ is an effective temperature which will be of central importance in our subsequent discussion.

\section{Analytical solution of the kinetic equation}\label{kinetical solution}
In the following, we discuss the effect of three different cross sections, $\sigma_0$, $\sigma_1$ and
$\sigma_2$, whose explicit functional dependence on relative velocity, together with that of their collisional frequencies, is respectively,
\begin{eqnarray}
\sigma_0(v)&=&\alpha_0 v^{-1}\cr\nu_0(v)&=&n\alpha_0\; ,\label{sigma0}
\\
\sigma_1(v)&=&\alpha_1\cr\nu_1(v)&=&n\alpha_1 v\; ,\label{sigma1}
\\
\sigma_2(v)&=&\alpha_2 v\cr\nu_2(v)&=&n\alpha_2 v^2\; ,\label{sigma2}
\end{eqnarray}
where $\alpha_0$, $\alpha_1$ and $\alpha_2$ are dimensional constants. We shall discuss in Section~\ref{collisional cross section} the physical meaning of the previous cross sections in dense astrophysical plasmas.

We state the hypothesis of absence of interference between the three collision types, namely we
assume that the total collisional frequency $\nu$ could be cast in the following approximate fashion,
\begin{displaymath}
\nu^2=\nu_0^2+\nu_1^2+\nu_2^2\; ,
\end{displaymath}
because different types of collision act significantly only in separate velocity intervals as evident from the functional dependencies reported in Eqs.~\ref{sigma0}, \ref{sigma1} and \ref{sigma2}.

Let us express now the solution of Eq.~\ref{general solution} as
\begin{equation}
f(v)=f(0)\cdot\exp[-I(v)]\; ,\label{general formula}
\end{equation}
where we have defined the integral function
\begin{eqnarray}
I(v)&=&\int_0^v \frac{\mu v'\textrm{d} v'}{\kb T\pm\frac{2}{3}\frac{F^2}{\kappa\mu n^2\alpha_0^2}\frac{1}{1+c_1 v'^2+c_2 v'^4}}\nonumber\\
& &=\frac{\mu}{\kb T}\int_0^v \frac{v'{\rm\, d}v'}{1+\tau\frac{1}{1+c_1 v'^2+c_2 v'^4}}\; ,\label{integral
function}
\end{eqnarray}
with $c_1=(\alpha_1/\alpha_0)^2$, $c_2=(\alpha_2/\alpha_0)^2$ and $\tau=\teff/T-1$, according to
Eqs.~\ref{effective temperature}, \ref{sigma0}, \ref{sigma1} and~\ref{sigma2}.

From Eq.~\ref{integral function}, we immediately obtain
\begin{equation}
I(v)=\frac{\mu v^2}{2\kb T}-\frac{\mu}{2\kb T}\tau I_1(v)\; ,\label{integral I}
\end{equation}
where
\begin{equation}
I_1(v)=\int_{0}^{v^2}\textrm{d} u\frac{1}{c_2 u^2+c_1 u+\tau+1}\; .\label{integral I1}
\end{equation}

Let us define now the following parameter,
\begin{displaymath}
K=-\frac{c_1^2}{4c_2}+\tau+1\; ,
\end{displaymath}
whose sign is physically relevant as we are about to show.

If $K<0$, Eq.~\ref{integral I1} gives, apart from an unimportant numerical term,
\begin{displaymath}
I_1=\frac{1}{2\sqrt{|K| c_2}}\left[\ln\left(\frac{2 c_2 v^2+ c_1 - 2\sqrt{|K| c_2}}{2 c_2 v^2+ c_1 + 2\sqrt{|K|
c_2}}\right) + \textrm{const}\right]\; ,
\end{displaymath}
which in turn, through Eqs.~\ref{integral I} and~\ref{general formula}, yields our first result (for $K<0$)
\begin{eqnarray}
f(v)&\propto&\exp{\left(-\frac{\mu v^2}{2\kb T}\right)}\nonumber\\
& &\times\left(\frac{2 c_2 v^2+ c_1 - 2\sqrt{|K| c_2}}{2 c_2 v^2+
c_1 + 2\sqrt{|K| c_2}}\right)^{\frac{\mu\tau}{4\kb T\sqrt{|K| c_2}}}\nonumber .
\end{eqnarray}

In this case, the cross section $\sigma_1$ dominates and the generalized random force field $F$ does not play any role in the region of interest for fusion reaction rate calculations in astrophysical plasmas, as the perturbation from the Maxwellian function vanishes in the limit $v\rightarrow +\infty$ (nevertheless, it can be of interest in studies of some atomic processes like radiative recombination, whose cross section increases as $v$ goes to zero and that, therefore, has rates sensibly modified by slight corrections at low velocity).

As far as astrophysical plasmas are concerned, a more interesting physical situation occurs when $K>0$, namely, if the condition
\begin{equation}
\frac{\teff}{T}>\frac{\nu_1^4}{4\nu_0^2\nu_2^2}=\frac{\alpha_1^4}{4\alpha_0^2\alpha_2^2}\; ,\label{physical
condition}
\end{equation}
holds. Condition~\ref{physical condition} is fulfilled providing the force field $F$ is
\begin{equation}
F>n\sqrt{\frac{3}{2}\kappa\mu\kb T\left(\frac{\alpha_1^4-4\alpha_0^2\alpha_2^2}{\alpha_2^2}\right)}\; ,\label{condition super}
\end{equation}
in case of super-diffusivity, or instead
\begin{equation}
F<n\sqrt{\frac{3}{2}\kappa\mu\kb T\left(\frac{4\alpha_0^2\alpha_2^2-\alpha_1^4}{\alpha_2^2}\right)}\; ,\label{condition sub}
\end{equation}
when considering sub-diffusivity.

From Eq.~\ref{integral I1} we obtain
\begin{eqnarray}
I_1(v)&=&\frac{1}{K}\int_{0}^{v^2}\textrm{d} u\left[\left(\sqrt{\frac{c_2}{K}}u+\frac{c_1}{2\sqrt{K c_2}}
\right)^2+1\right]^{-1}\nonumber\\
& & =\frac{1}{\sqrt{K c_2}}\left[\arctan\left(\sqrt{\frac{c_2}{K}}v^2+\frac{c_1}{2\sqrt{K c_2}}\right)\right.\nonumber\\
& &\left.-\arctan\frac{c_1}{2\sqrt{K c_2}}\right]\; .\label{important solution}
\end{eqnarray}

Starting from Eqs.~\ref{integral I}, \ref{integral I1} and~\ref{important solution} we can express $I(v)$ as a formal series of powers of $v^2$,
\begin{displaymath}
I(v)=\frac{\mu v^2}{2\kb\teff}+\delta\left(\frac{\mu v^2}{2\kb\teff}\right)^2+\gamma\left(\frac{\mu
v^2}{2\kb\teff}\right)^3+\cdots\; ,
\end{displaymath}
where
\begin{displaymath}
\delta=\pm\frac{2}{3}\frac{F^2}{\kappa\mu^2 n^2}\frac{\alpha_1^2}{\alpha_0^4}\; ,
\end{displaymath}
and
\begin{displaymath}
\gamma=\pm\frac{8}{9}\frac{F^2 \kb T}{\kappa\mu^3 n^2}\frac{\alpha_2^2}{\alpha_0^4}
\left(1-\frac{\alpha_1^4}{\alpha_0^2\alpha_2^2}\right)+\frac{16}{27}\frac{F^4}{\kappa^2 \mu^4 n^4}
\frac{\alpha_2^2}{\alpha_0^6}\; ,
\end{displaymath}
both being $|\delta|,|\gamma|\ll 1$.

Therefore, the final form of the one-body distribution function under generalized random fields reads
\begin{eqnarray}
f(v)&\propto&\exp\left[-\frac{\mu v^2}{2\kb\teff}\right]\nonumber\\
& &\times\exp\left[-\delta\left(\frac{\mu v^2}{2\kb\teff}\right)^2\right]\exp\left[-\gamma\left(\frac{\mu v^2}{2\kb\teff}\right)^3\right] ,\nonumber
\end{eqnarray}
that corresponds to an energy probability factor
\begin{eqnarray}
f(\ep)&\propto&\exp\left[-\frac{\ep}{\kb\teff}\right]\nonumber\\
& &\times\exp\left[-\delta\left(\frac{\ep}{\kb\teff}\right)^2\right]
\exp\left[-\gamma\left(\frac{\ep}{\kb\teff}\right)^3\right] ,\nonumber\\ \label{final solution}
\end{eqnarray}
where $\ep=p^2/(2\mu)$ is the centre-of-mass kinetic energy, given the linear momentum ${\bf p}=\mu {\bf v}$.

It is noteworthy that our result in Eq.~\ref{final solution} may be related at least for small deformations to the non-extensive distribution function at the same effective temperature $\teff$~\cite{nonextensive}
\begin{equation}
f(\ep)\propto\left[1-(1-q)\frac{\ep}{\kb\teff}\right]^{\frac{1}{1-q}}\; ,\label{nonext distribution}
\end{equation}
where $q$ is called the entropic parameter. As can be straightforwardly shown after some manipulations, in the low deformation limit $(q-1)\ep/(\kb\teff)\rightarrow 0$, Eq.~\ref{nonext distribution} reduces to Eq.~\ref{final solution}, provided that $\delta=(1-q)/2$. Thus, this condition establish a link between the entropic parameter $q$ and our parameter $\delta$ that comes from microfield strength and cross sections. We point out that, in the same limit, also other distributions of generalized statistics have the same behaviour, as explained in Ref.~\cite{Beck}.

We recall that in recent past we have shown that, if the generalized random force is due to an electric microfield
distribution, the parameter $\delta$ can be related to the non-extensive (Tsallis) entropic parameter $q$ and the
following analytical expression can be derived,
\begin{displaymath}
\delta=\frac{1-q}{2}=12\Gamma^2\alpha^4\; ,\label{delta-q}
\end{displaymath}
where $\Gamma$ is the plasma parameter and $\alpha$ is a dimensionless parameter accounting for ion correlations in the ion-sphere model ($0.4<\alpha<1$)~\cite{CKLLMQ1999}.

From Eq.~\ref{final solution} it follows that there exist three different intervals of relative velocity in
which the evaluated corrections due to the random field are sufficiently large to be noteworthy. First of all, we
observe that if $\ep\sim\kb\teff$, the dominant factor is $\exp[-\ep/(\kb\teff)]$, namely the Maxwellian factor
characterized by the cross section $\sigma_0$. The exponential factor with the $\delta$ parameter, corresponding to the collisional cross section $\sigma_1$, becomes not negligible with respect to the Maxwellian only when
$\ep\sim\kb\teff/|\delta|$; it is also often called the Druyvenstein factor. Finally, the third term $\exp[-\gamma(\ep/\kb\teff)^3]$ arises when $\ep\sim |\delta/\gamma|\kb\teff$; as we shall sketch in Section~\ref{sect quantum uncertainty}, it can be related with quantum energy-momentum uncertainty effects in dense astrophysical plasmas. 

In conclusion of this section, let us summarize that if the random force field is absent or negligible, in spite of the presence of whatever kind of collisional cross sections, all stationary states which are solutions of the kinetic equation have an analytical expression that coincides with the equilibrium Maxwellian distribution function. Therefore, the non-extensive statistical description, at least in a classical framework, requires as a first condition that particles be subjected to a sufficiently strong random force field and, as second condition, that a constant collisional cross section (or depending on higher positive powers of velocity) should act among the particles of the system.

\section{Collisional cross sections in astrophysical plasmas}\label{collisional cross section}
We want to discuss the physical meaning of the collisional processes related to $\sigma_0$, $\sigma_1$ and $\sigma_2$, defined in Section~\ref{kinetical solution} and inserted into the kinetic equation in order to derive the one-body distribution function.

The cross section $\sigma_0(v)$, defined in Eq.~\ref{sigma0}, is the most important one as it originates the
well-known Maxwellian distribution function even in presence of a generalized random field. That the solution of
the kinetic equation (Eq.~\ref{kinetic equation}), at first order, be the Maxwellian function is our first unavoidable requirement, because the actual kinetic solution for $F=0$ is indeed the Maxwellian, and we are dealing only with slight corrections.

Following Ref.~\cite{Present} we can state that, starting from an interaction force that depends on distance as $r^{-s}$, the corresponding cross section is $\sigma(v)\propto v^{-4/(s-1)}$. Therefore, in the case of the cross section $\sigma_0(v)\propto v^{-1}$, the underlying force goes like $r^{-5}$, while the potential energy is proportional to $r^{-4}$, and we can interpret it as the cross section due to the interaction between an ionic charge and an electric dipole induced by the ion on the neutral system of charges composing a Debye sphere~\cite{Golant}. 

On the contrary, if we considered a pure Coulomb interaction due to a force $F_C\propto r^{-2}$ (with $s=2$), it would give a collisional cross section proportional to $v^{-4}$; however, this case is not physically suitable in presence of an intensive random force field because of induced divergences in the distribution function at small velocities. Krook and Wu showed in the past~\cite{Wu} that collisional cross sections going like $v^{-1}$ and $v^{-3}$ always produce, after a sufficient long time, a Maxwellian distribution; however their system is not subjected to a random force field.

The cross section $\sigma_1(v)$, defined in Eq.~\ref{sigma1}, has been introduced by Ichimaru~\cite{Ichimaru1992} in the framework of an ion-sphere model for non-ideal and weakly-coupled plasmas with a $\Gamma$ parameter of order unity, and with a small number of ions inside the Debye sphere. In these physical conditions the collisional cross section, directly derived from the pure Coulomb one, is constant according to the approximations of the model and it can be cast in the simple form
\begin{displaymath}
\sigma_1 (v)\approx 2\pi (\alpha a)^2\; ,
\end{displaymath}
where $\alpha$ is the correlation factor already introduced in Section~\ref{delta-q} and $a$ is the interparticle distance. The correction due to $\sigma_1$ on the probability function of energy is of order $\exp[-\delta\ep^2/(\kb\teff)^2]$, and shows the same behaviour of the so called non-extensive corrective factor (see Ref.~\cite{non-extensivity}). 

The cross section $\sigma_2(v)$ will be described in the next section.

\section{Effects of quantum energy-momentum uncertainty on the equilibrium distribution function}\label{sect quantum uncertainty}
Here we introduce simple arguments to synthetically explain the meaning and justify the use of the cross section $\sigma_2(v)$ defined in Eq.~\ref{sigma2} and, at the same time,  to show a possible link between quantum energy-momentum uncertainty and non-extensivity.

Quantum energy-momentum uncertainty in weakly non-ideal dense stellar plasmas influences thermonuclear rates. In fact, in classical physics, energy $\varepsilon$ and momentum ${\bf p}$ (or $\ep={\bf p}^2/2\mu$, with $\mu$ reduced mass) are linked by the dispersion relation $\delta(\varepsilon,\ep)=\delta(\varepsilon-\ep)$. Nevertheless, if the particles cannot be considered free, $\varepsilon$ and $\ep$ are independent variables. For instance, an ion tunnelling the Coulomb barrier during a thermonuclear fusion reaction can interact simultaneously with many other particles. In this case, the dispersion relation is given by the function $\delta_{\gamma}(\varepsilon,\ep)$ defined using the ansatz of Kadanoff and Baym~\cite{BaymKadanoff}. Under equilibrium conditions, and this time without any random force field, the energy and momentum generalized distribution function can be
written as
\begin{displaymath}
f(\varepsilon,\ep)=\frac{n(\varepsilon)}{\pi}\delta_{\gamma}(\varepsilon,\ep)\; ,
\end{displaymath}
with
\begin{displaymath}
\delta_{\gamma}(\varepsilon,\ep)=\frac{\mathrm{Im}
\Sigma^R(\varepsilon,\ep)}{(\varepsilon-\ep-\mathrm{Re}\Sigma^R)^2+(\mathrm{Im}\Sigma^R)^2}\; ,
\end{displaymath}
where $n(\varepsilon)$ is the particle number distribution and $\Sigma^R(\varepsilon,\ep)$ is the mass operator
for the one-particle retarded Green function.

Galitski\v{\i} and Yakimets~\cite{Galitskii} (see also Refs.~\cite{quantum,Starostin}) derived that the quantum energy-momentum indeterminacy and a non-zero value of $\mathrm{Im}\Sigma^R$ lead to the non-exponential tail of the energy probability factor $f(\ep)$.

We limit ourselves to the case of a dispersion relation of Lorentz type,
\begin{displaymath}
\delta_{\gamma}(\varepsilon,\ep)=\frac{1}{\pi}\frac{\gamma(\varepsilon,\ep)}
{(\varepsilon-\ep)^2+\gamma^2(\varepsilon,\ep)}\; ,
\end{displaymath}
with
\begin{displaymath}
\gamma(\varepsilon,\ep)=\hbar \nu_{coll}^T (\varepsilon,\ep)=\hbar
n\sigma_t(\ep)\left(\frac{2\varepsilon}{\mu}\right)^{1/2}
\end{displaymath}
where $\nu_{coll}^T (\varepsilon,\ep)$ is the total collision frequency and $\sigma_t(\ep)$ is the collisional
cross section.

Let us make the example of a pure Coulomb interaction. We have that
\begin{displaymath}
\gamma(\varepsilon,\ep)=\frac{2\pi\hbar n\mathrm{e}^4}{\ep^2}\left(\frac{2\varepsilon}{\mu}\right)^{1/2}\; ,
\end{displaymath}
and the momentum distribution becomes
\begin{eqnarray}
f(\ep)&=&\int\textrm{d}\varepsilon f(\varepsilon,\ep)=\int\textrm{d}\varepsilon\frac{n(\varepsilon)}{\pi}\delta_{\gamma}(\varepsilon,\ep)\nonumber\\
& &=\frac{2}{\pi^{3/2}}\frac{\sqrt{\ep}}{(\kb T)^{3/2}}\left[f_{MB}(\ep)+f_Q(\ep)\right]\, ,\label{general distribution}
\end{eqnarray}
with
\begin{displaymath}
f_{MB}(\ep)=\exp\left(-\frac{\ep}{\kb T}\right)\; ,
\end{displaymath}
and
\begin{displaymath}
f_Q(\ep)=\frac{2}{3\pi}C (\kb T)^{3/2}\frac{1}{\ep^4}\; ,
\end{displaymath}
where $C$ is a constant depending on the density $n$.

At high momenta the last term in Eq.~\ref{general distribution} can be many orders of magnitude
greater than the first one and represents an enhancement of the tail, with important consequences in the
calculations of nuclear fusion rates.

We want to verify if, by using a certain elastic collision cross section, we could obtain from the quantum uncertainty effect the
non-extensive Tsallis distribution, limiting ourselves, for simplicity, to the case of entropic parameter
$q>1$~\cite{nonextensive}.

Following (and adapting to our present needs energy fluctuations instead of temperature fluctuations) the
approach outlined by Beck and Cohen~\cite{Beck}, we may state that any non-Maxwellian energy probability function
should be cast in the form of a Laplace transformation of the function $\delta_f(E,\ep)$ which describes the non ideality of the system~\cite{Cortona},
\begin{displaymath}
f(\ep)=\int_0^{+\infty}\mathrm{d} E \exp\left(-\frac{E}{\kb T}\right)\cdot\delta_f(E,\ep)\; .
\end{displaymath}

The function $\delta_f(E,\ep)$ must be assumed to be a Gamma- (or $\chi^2$-) function, in order to obtain that $f(\ep)$ be a non-extensive (Tsallis) distribution~\cite{Beck2001,Beck}.

Let us compare this integral with the integral of Eq.~\ref{general distribution}, that can be written explicitly
as $f(\ep)=\int\mathrm{d}\varepsilon\exp(-\varepsilon/\kb T)\delta_{\gamma}(\varepsilon,\ep)$. Quantum uncertainty and non-extensivity are two different and distinct causes of deformation of the Maxwell-Boltzmann distribution.
Nevertheless they give the same effect if the microscopic interaction among the particles (i.e. the collisional cross section) is of a particular nature as we discuss below.

Let us pose attention to the physical property of interest is the `width' $D_f$ of the $\delta_f (E,\ep)$ distribution; it can be shown that for the non-extensive distribution function, the width is $D_{NE}\sim\ep^2$, while for the quantum uncertainty the $D_Q\sim\sigma^2(\ep)\ep$ relation holds, where $\sigma(\ep)$ is the collisional cross section. If we now impose that super-extensivity and quantum uncertainty give the same physical effect on distribution functions, we should require that $\sigma(\ep)\propto\sqrt{\ep}$ or, in terms of relative velocity, $\sigma(v)=\sigma_2(v)\propto v$. Thus, the cross section $\sigma_2$ that we have used in Section~\ref{kinetical solution}, is strongly related with both quantum and non-extensive statistical effects. The non-extensive and the Galitski\v{\i}-Yakimets distributions result given by the same expression.

Let us recover the behaviour of the interaction force responsible of the cross section
$\sigma(\ep)\sim\sqrt{\ep}$. We can write its dependence on the relative coordinate $r$ of the two interacting
particles as~\cite{Present}
\begin{displaymath}
F(r)=f_0\left(\frac{r}{R_0}\right)^{-s}\; ,
\end{displaymath}
where $f_0$ is a dimensional constant, $R_0$ is a characteristic distance of the two-body center of mass with respect to a given origin and $s$ is a negative or positive integer.

Defining the collisional cross section as
\begin{displaymath}
\sigma=\pi d^2\; ,
\end{displaymath}
with
\begin{displaymath}
d\sim\left(f_0\frac{\mu}{|{\bf p}|^2}\right)^{1/(s-1)}\; ,
\end{displaymath}
in order to have the requested behaviour of $\sigma(\ep)\propto\sqrt{\ep}$, we must set $s=-3$. Let us recall that the case $s=-3$ is, from the point of view of the orbit differential equation of motion, one of the integrable cases, with solutions given in terms of elliptical functions~\cite{Goldstein}.

Therefore the interacting force responsible of the collisions that lead to $\sigma(\ep)\sim\sqrt{\ep}$ reads
\begin{displaymath}
F_Q(r)=\cases{f_{Q_0}\left(\frac{r}{R_0}\right)^3 & $r\le R_0$ \cr 0 & $r>R_0$}
\end{displaymath}
where the cut-off is needed in order to avoid divergences of the potential energy.

We may argue that the force $F_Q(r)$ can be understood as a tidal-like force~\cite{Lamb} if we assume that an attractive central force of intensity $f_{Q_0}$, centered at a distance $R_0$ from the center of mass of the two interacting particles separated by a distance $r$, is superimposed. The tidal-like force acts globally over all the particles of the system. This is the dynamical requirement to recover the non-extensive (Tsallis) distribution in the framework of quantum energy-momentum uncertainty. It is worth and peculiar to remark that by applying the virial theorem to this case we obtain a negative kinetic energy, which is, in fact, understandable and admissible by the uncertainty principle.

We derive the analytical expression of $q$ by equalling the complete expressions of $D_{NE}$ and $D_{Q}$. We obtain
\begin{displaymath}
q=1+\frac{(\hbar c)^2 n^2(\Sigma_l)^2}{2^5 (\mu c^2)}\frac{R_0^3}{f_{Q_0}}\; ,
\end{displaymath}
where $n$ is the plasma density and $\Sigma_l$ is of order of unity~\cite{Starostin}.

The correction to the unity can be thought due to the many-body effect over the two-body interacting system. As an example, let us make the following numerical approximate evaluation of $f_{Q_0}$: if the correction is of order of ten per cent, the density $n\approx 10^{-14}\,\mathrm{fm}^{-3}$, and $R_0\approx 10^5\,\mathrm{fm}$, we obtain, for a proton plasma ($\mu c^2\approx 460\,\mathrm{MeV}$),
\begin{displaymath}
f_{Q_0}\approx 10^{-12}\,\mathrm{MeV/fm}\; .
\end{displaymath}

Before concluding this section we remark that the non-extensive distribution usually describes metastable states or
stationary states of non-equilibrium systems. On the contrary, in this case, quantum uncertainty with collisional
cross section $\sigma(\ep)\sim\sqrt{\ep}$ gives a distribution function which belongs to an equilibrium state,
although different from the Maxwell-Boltzmann distribution. Other generalized distributions have been recently proposed~\cite{Kaniadakis}. For situations with small deformation our argument are valid also for these distributions.

\section{Conclusions}
We have set a kinetic equation suitable to describe the stationary states of a weakly non-ideal plasma of a stellar core subject to generalized random forces. Provided that a random force satisfying condition~\ref{condition super} for super-extensivity or~\ref{condition sub} for sub-extensivity is present, the momentum distribution function can be cast in the simple fashion of Eq.~\ref{final solution} in which, besides the well-known Maxwellian factor, other terms also contribute.

The momentum distribution function is formally identical to the non-extensive distribution (when $q>1$) expanded in
powers of $(1-q)$ for small deformation. An analytical expression of $q$, the entropic parameter, can be derived in terms of the elastic collision cross sections acting among the particles of the system.

The main point is that each correction factor is due to a particular collisional process between ions, and that
each of them contributes in a well defined interval of relative velocity, as shown at the end of
Section~\ref{kinetical solution}. All these corrections are small, nevertheless they are not negligible at high energy, i.e. in the region of ion spectrum of predominant interest for calculations of nuclear reaction rates in astrophysical plasmas.

We have stressed that in physical conditions as, for example, stars with $\Gamma\gtrsim 1$, many
collisional processes may be active, maybe at the same time, and that each one of them is described by a cross
section with a dependence over velocity stronger (proportional to $v^{-1}$, $v^0$ or even $\propto v$) than the
simple Coulomb scattering (proportional to $v^{-4}$). This fact is intimately related to statistical many-body
effects and represents a link between dynamics (the type of two-body elastic collisions) and statistical mechanics (the momentum distribution function of the stationary states involved).

Finally, in the framework of a quantum many-body description of the equilibrium state, considering the
energy-momentum uncertainty due to the non-commutativity of position and momentum operators, we have found that if
the collisional cross section $\sigma(\ep)$ behaves like $\sqrt{\ep}$, the distribution function coincides with the
non-extensive (Tsallis) distribution function with $q>1$. The requested behaviour of the cross section
$\sigma(\ep)$ is due to an interaction similar to a tidal-like force. Therefore the analogy
between quantum uncertainty effect on the distribution and non-extensive effect is achieved provided that an
overall attractive interaction is superimposed to the two-body interaction. This represents again a possible link between dynamics and statistical mechanics.

\end{document}